\documentclass[conference]{IEEEtran}
\IEEEoverridecommandlockouts
\usepackage{cite}
\usepackage{amsmath,amssymb,amsfonts}
\usepackage{algorithmic}
\usepackage{graphicx}
\usepackage{textcomp}
\usepackage{xcolor}
\usepackage{float}
\usepackage{subfigure}
\usepackage{hyperref}
\usepackage[misc]{ifsym}
\usepackage[ruled,linesnumbered]{algorithm2e}
\def\BibTeX{{\rm B\kern-.05em{\sc i\kern-.025em b}\kern-.08em
    T\kern-.1667em\lower.7ex\hbox{E}\kern-.125emX}}
\newcommand{\tabincell}[2]{\begin{tabular}{@{}#1@{}}#2\end{tabular}}
\begin{document}

\title{MMD-MIX: Value Function Factorisation with Maximum Mean Discrepancy for Cooperative Multi-Agent Reinforcement Learning
}

\author{\IEEEauthorblockN{Zhiwei Xu, Dapeng Li, Yunpeng Bai, Guoliang Fan
}
\IEEEauthorblockA{\textit{Fusion Innovation Center, Institute of Automation, Chinese Academy of Sciences}}
\textit{School of Artificial Intelligence, University of Chinese Academy of Sciences}\\
Beijing, China\\
xuzhiwei2019@ia.ac.cn, lidapeng2020@ia.ac.cn, baiyunpeng2020@ia.ac.cn, {guoliang.fan@ia.ac.cn}
\\
}


\maketitle

\begin{abstract}
In the real world, many tasks require multiple agents to cooperate with each other under the condition of local observations. To solve such problems, many multi-agent reinforcement learning methods based on Centralized Training with Decentralized Execution have been proposed. One representative class of work is value decomposition, which decomposes the global joint Q-value $Q_\text{jt}$ into individual Q-values $Q_a$ to guide individuals’ behaviors, e.g. VDN (Value-Decomposition Networks) and QMIX. However, these baselines often ignore the randomness in the situation. We propose MMD-MIX, a method that combines distributional reinforcement learning and value decomposition to alleviate the above weaknesses. Besides, to improve data sampling efficiency, we were inspired by REM (Random Ensemble Mixture) which is a robust RL algorithm to explicitly introduce randomness into the MMD-MIX. The experiments demonstrate that MMD-MIX outperforms prior baselines in the StarCraft Multi-Agent Challenge (SMAC) environment.
\end{abstract}


\section{Introduction}
Recently, deep reinforcement learning (DRL), as a combination of deep learning (DL) and reinforcement learning (RL), plays an increasingly important role in sequential decision-making problems. In reinforcement learning, the agent interacts with the environment in real-time based on the action modes, current states and the corresponding feedback. The agent aims to find a so-called policy to predict and maximize the expected return\cite{Sutton2005ReinforcementLA}. This kind of thinking makes reinforcement learning have significant effects in many application scenarios, such as Atari games \cite{Mnih2013PlayingAW}, robot control\cite{OpenAI2019SolvingRC}, and autonomous vehicles\cite{Faust2018PRMRLLR}.

In order to get better results and a more stable learning process, a series of reinforcement learning algorithms with distributional perspectives have been proposed. Intuitively speaking, the main difference from general reinforcement learning algorithms is that the expected return predicted by these distributional algorithms is distribution rather than a single value. Categorical-DQN (C51)\cite{Bellemare2017ADP}, as the first proposed distributional reinforcement learning algorithm, divides the range of possible return into a few bins and uses a neural network to estimate the probability of each one. In QR-DQN\cite{Dabney2018DistributionalRL} and its related variants IQN\cite{Dabney2018ImplicitQN} and FQF\cite{Yang2019FullyPQ}, the return quantiles on different quantile fractions will be computed, and the Huber quantile regression loss between Bellman updated distribution and the current actual return distribution will be minimized according to quantile regression. However, MMD-DQN\cite{Nguyen2020DistributionalRL} is different from the two methods described above in that it is not limited to any predefined statistical functions. MMD-DQN implicitly expresses the expected distribution in the form of a set of particles. By minimizing the maximum mean discrepancy (MMD) distance between the two sets of particles, reducing the distance between the Bellman updated distribution and the current distribution can be achieved. In this way, MMD-DQN can approximate the return distribution more flexibly. In addition, \cite{Agarwal2020AnOP} recently proposed a robust ensemble Q-Learning algorithm REM (Random Ensemble Mixture), which enforces optimal Bellman consistency on random convex combinations of multiple Q-value estimates and can be comparable to distributional reinforcement learning algorithms in off-line situations. However, if the above-mentioned single-agent reinforcement learning algorithms are directly applied to multi-agent problems, it will lead to new problems: the environment becomes nonstationary from the point of view of each agent.

Besides, multi-agent cooperative tasks, as a special case of multi-agent reinforcement learning, have quickly attracted people's attention in recent years. Because of the partial observation limitation, the existence of other agents introduces environmental instability factors. This is disastrous for the learning of the agent. To solve this problem, in addition to adding to the communication between agents\cite{Sukhbaatar2016LearningMC, Foerster2016LearningTC, Peng2017MultiagentBN, Jiang2018LearningAC, Kim2019LearningTS} or using a "decentralized actor, centralized critic" training approach\cite{Lowe2017MultiAgentAF, Foerster2018CounterfactualMP, Iqbal2019ActorAttentionCriticFM}, there is also a representative method, which is to decompose the joint value function. VDN (Value-Decomposition Networks)\cite{Sunehag2018ValueDecompositionNF} obtains the joint action-value which is a linear sum of individual action-values and optimizes the individual action-values of all agents by optimizing the joint action-value. Based on monotonicity, QMIX\cite{Rashid2018QMIXMV} uses a set of hypernetworks\cite{Ha2017HyperNetworks} to generate a mixing net to non-linearly approximate the joint action-value. By further extending additivity and monotonicity to factorizability, QTRAN\cite{Son2019QTRANLT} extends the reward type used by value decomposition, but it also slightly increases the computational complexity. Generically, value decomposition methods ignore the randomness because of modeling the mean value of the joint state-action value.

\noindent
\textbf{Contribution:}In this paper, our contributions are as follows:
\begin{itemize}
\item We propose MMD-MIX, which combines distributional reinforcement learning with value decomposition methods to improve the ability to adapt to randomness.

\item Depending on the characteristics of REM, we weight and sum the particles output by MMD-MIX to obtain a new set of particles, making each new particle a random convex combination of the original particles. Through this method of introducing random noise, the robustness and exploration ability of the algorithm are improved. 

\item We prove that the proposed algorithm is better than the previous baselines through the experimental results on SMAC\cite{samvelyan19smac}.

\end{itemize}
\section{Background}

\subsection{Dec-POMDP}
Dec-POMDP\cite{Oliehoek2016ACI} is a fully cooperative multi-agent task, which can be represented by the tuple $G = (\mathcal{S}, \mathcal{U}, P, r, \mathcal{Z}, O, n, \gamma)$. $s \in \mathcal{S}$ represents the state of the environment. Each agent $a \in \mathcal{A} \equiv \{1, \dots,n\}$ will output the action $u_a \in \mathcal{U}$ when interacting with the environment. The actions of all agents at the same time are combined into a joint action $\boldsymbol{u} \in \boldsymbol{\mathcal{U}} \equiv \mathcal{U}^n$. $P$ represents a state transition function, and we can obtain the next state of the environment according to $P\left(s^{\prime} \mid s, \boldsymbol{u}\right): \mathcal{S} \times \boldsymbol{\mathcal{U}} \times \mathcal{S} \rightarrow[0, 1]$. Different from the ones in other multi-agent reinforcement learning tasks, the agents in Dec-POMDP share the same reward function : $r(s, \boldsymbol{u}): \mathcal{S} \times \boldsymbol{\mathcal{U}} \rightarrow \mathbb{R}$. The individual observation of each agent is represented by $z \in \mathcal{Z}$, and $z$ can be obtained by observation function $O(s, a): \mathcal{S} \times \mathcal{A} \rightarrow \mathcal{Z}$. $\gamma$ is the discount factor. The goal of the Dec-POMDP scenario is to maximize the discounted return $R^{t}=\sum_{i=0}^{\infty} (\gamma)^{i} r^{t+i}$.

In the partially observable scenario, historical information can be presented to make the learning process more stable. The policy of each agent $a$ can be expressed as $\pi_{a}\left(u_{a} \mid \tau_{a}\right): T \times \mathcal{U} \rightarrow [0, 1] $, where $\tau_{a} \in T \equiv(\mathcal{Z} \times \mathcal{U})$ is the action-observation history of the agent $a$. The joint policy $\pi$ of all agents corresponds to a joint action-value function: $Q^{\pi}\left(s^{t}, \boldsymbol{u}^{t}\right)=\mathbb{E }_{s^{t+1: \infty}, \boldsymbol{u}^{t+1: \infty}}\left[R^{t} \mid s^{t}, \boldsymbol{u}^{t }\right]$.
\subsection{Distributional RL via maximum mean discrepancy}
\subsubsection{Maximum mean discrepancy}
MMD is often used in two-sample tests in statistics to distinguish two distributions by finite samples\cite{Gretton2012AKT, Gretton2006AKM}. Assuming that $\mathcal{X}$ is a non-empty compact metric space, $\mathbb{P}$ and $\mathbb{Q}$ are two probability measures on $\mathcal{X}$, $X$ and $Y$ are two random variables corresponding to the distributions $\mathbb{P}$ and $\mathbb{Q}$. $\mathcal{F}$ is a family of functions: $f: \mathcal{X} \rightarrow \mathbb{R}$, and $\mathcal{F}$ is chosen to be a unit ball in a reproducing kernel Hilbert space (RKHS) $\mathcal{H}$ associated with a kernel $k(\cdot, \cdot)$. Then the MMD distance between $\mathbb{P}$ and $\mathbb{Q}$ can be noted as:
\begin{equation}
    \label{eq:mmd_distance}
    \begin{split}
        &\operatorname{MMD}^{2}(\mathbb{P}, \mathbb{Q} ; \mathcal{F})=
        \left\|\psi_{\mathbb{P}}-\psi_{\mathbb{Q}}\right\|_{\mathcal{H}}^{2}=\\
        &\mathbb{E}\left[k\left(X, X^{\prime}\right)\right]
        +\mathbb{E}\left[k\left(Y, Y^{\prime}\right)\right]-2 \mathbb{E}[k(X, Y)],
    \end{split}
\end{equation}
where Bochner integral $\psi_\mathbb{P} := \int_{\mathcal{X}} k(x, \cdot) \mathbb{P}(dx)$ is the mean embedding of $\mathbb{P}$ into $\mathcal{H}$. If $\psi_\mathbb{P}$ is injective (for example, when $k(\cdot ,\cdot)$ is Gaussian or Laplacian kernel), MMD is a measure of $\mathcal{P}(\mathcal{X})$. $X^\prime$ and $Y^\prime$ are two random variables in the distributions $\mathbb{P}$ and $\mathbb{Q}$, and they are independent of $X$ and $Y$ respectively.

Given empirical samples $X=\left\{x_{1}, \cdots, x_{M}\right\} \sim \mathbb{P}$ and $Y=\left\{y_{1}, \cdots, y_{N}\right\} \sim \mathbb{Q}$, we can get the squared MMD distance from Equation (\ref{eq:mmd_distance}):

\begin{equation}
    \label{eq:square_mmd}
    \begin{split}
        &\mathcal{L}_{M M D^{2}}\left(\left(x_{i}\right),\left(y_{i}\right) ; k\right)=\\
        &\frac{1}{N^{2}} \sum_{i, j} k\left(x_{i}, x_{j}\right) 
        +\frac{1}{M^{2}} \sum_{i, j} k\left(y_{i}, y_{j}\right)-\frac{2}{N M} \sum_{i, j} k\left(x_{i}, y_{j}\right).
    \end{split}
\end{equation}
Many works have achieved better results by minimizing the MMD distance to fit the specific distribution, e.g. GMMN\cite{Li2015GenerativeMM} and MMD-GAN\cite{Li2017MMDGT}.

\subsubsection{MMD-DQN}
Instead of the original mean square error (MSE), MMD is incorporated into DQN to measure the distance between Bellman target distribution with current distribution. Similar to other distributional reinforcement learning methods, the neural network representing the action-value function outputs a distribution $Z$. In MMD-DQN, this distribution $Z$ is presented by $N$ particles. Suppose that given a trajectory sample $(s,u,r,s^\prime, u^{\prime})$, $s$ and $s^\prime$ represent the current state and next state of the environment respectively. Similarly, $u$ and $u^\prime$ represent the current action and next action of the agent. And $r$ represents the current reward. Then the current distribution can be expressed as $\left(Z_{\theta}(s, u)_{i}\right)_{i=1}^{N}$, the Bellman target distribution is 
\begin{equation*}
        \hat{\mathcal{T}} Z_{i}:=r+\gamma Z_{\theta^-}\left(s^{\prime}, u^{\prime}\right)_{i}, \;\;\forall i \in\{1, \ldots, N\}.
\end{equation*}
According to the Equation (\ref{eq:mmd_distance}), the square MMD distance can be obtained as 
\begin{equation}
\label{eq:loss}
\mathcal{L}_{MMD^{2}}\left(\left(Z_{\theta}(s,u)_{i}\right)_{i= 1}^{N},\left(\hat{\mathcal{T}} Z_{i}\right)_{i=1}^{N}; k\right).
\end{equation}
It should be noted that $k$ is a kernel function. General kernel functions, such as the unrectified triangle kernel $k(x, y)=-\|x-y\|^{p}$ or Gaussian kernel $k(x , y)=\exp \left(-\frac{1}{h}(x-y)^{2}\right)$ can be applied. Intuitively, MMD-DQN evolves particle $\left(Z_{\theta}(s, u)_{i}\right)_{i=1}^{N}$ by minimizing $\mathcal{L}_{M M D^{2}}$ given by Equation (\ref{eq:loss}).

\subsection{Random Ensemble Mixture}
REM is a method similar to ensemble reinforcement learning, which approximates the Q-values via an ensemble of parameterized Q-value functions. The key idea of REM is to obtain a new Q-value by a randomly convex combination of these Q-value estimates, i.e., the robustness of the algorithm is improved by training a set of Q-function approximators defined by mixing probabilities on a ($K-1$)-simplex.

REM is an extension on the basis of DQN. In the training process, for each mini-batch, a categorical distribution $\alpha \sim \mathbf{P}_\Delta$ is randomly drew to perform a weighted sum of multiple Q-value estimates output by the last layer of DQN, where $\mathbf{P}_\Delta$ represents a probability distribution over the standard ($K-1$)-simplex $\Delta^{\mathrm{K}-1}=\left\{\alpha \in \mathbb{R}^{\mathrm{K}}: \alpha_{1}+\alpha_{2}+\cdots+\right.\left.\alpha_{K}=1, \alpha_{k} \geq 0 , k=1, \ldots, K\right\}$. Temporal difference error can take the form of:
\begin{equation*}
    \begin{split}
        \sum_{k=1}^K \alpha_{k} Q_{\theta}^{k}(s, u)-r-\gamma \max _{u^{\prime}} \sum_{k=1}^K \alpha_{k} Q_{\theta^{\prime}}^{k}\left(s^{\prime}, u^{\prime}\right),
    \end{split}
\end{equation*}
where $\theta$ represents the parameters of the current Q network, and $\theta^\prime$ represents the parameters of the target Q network, and $K$ represents the output dimension of the last layer of DQN. When the assumptions in Proposition 1 proposed by \cite{Agarwal2020AnOP} is satisfied, convergence can be achieved by minimizing the mean square error or huber loss of temporal difference error.

Regarding the distribution $\mathbf{P}_\Delta$, a relatively simple distribution is often chosen. The common method is to draw a set of K values i.i.d. from Uniform (0, 1) and normalize them to get a valid categorical distribution, i.e., $\alpha_k$ can be obtained from the following formulas:
\begin{equation*}
    \alpha^\prime_k \sim U(0,1),\quad \quad \alpha_k = \alpha^\prime_k / \sum_{k=1}^K \alpha_k^\prime,
\end{equation*}
where $k \in \{1, 2,\dots, K\}$.

\subsection{Value-Decomposition Multi-Agent RL}
IGM (Individual-Global-Max) \cite{Son2019QTRANLT} defines the optimal consistency between an individual agent and all agents as a whole. Using $Q_{\text{jt}}$ and $Q_a$ to represent joint action-value function and individual action-value function, IGM can be expressed as:
\begin{equation*}
    \arg \max _{\boldsymbol{u}} Q_{\mathrm{jt}}(\boldsymbol{\tau}, \boldsymbol{u})=\left(\begin{array}{c}
\arg \max _{u_{1}} Q_{1}\left(\tau_{1}, u_{1}\right) \\
\vdots \\
\arg \max _{u_{n}} Q_{n}\left(\tau_{n}, u_{n}\right)
\end{array}\right),
\end{equation*}
where $\boldsymbol{\tau} \in T^n$ represents the joint action-observation histories of all agents.

Almost all algorithms based on value decomposition satisfy the IGM condition and finally achieve convergence. VDN fits the joint action-value function by adding up the individual action-value functions of all agents. It takes advantage of additivity:
\begin{equation*}
    Q_{\text {jt }}(\boldsymbol{\tau}, \boldsymbol{u})=\sum_{a=1}^{n} Q_{a}\left(\tau_{a}, u_{a}\right).
\end{equation*}

QMIX uses monotonicity to meet the IGM condition:
\begin{equation*}
    \frac{\partial Q_{\mathrm{jt}}(\boldsymbol{\tau}, \boldsymbol{u})}{\partial Q_{a}\left(\tau_{a}, u_{a}\right)} \geq 0, \quad \forall a \in \{1,\dots,n\}.
\end{equation*}
The structure of QMIX consists of three parts: agent networks, a mixing network, and a set of hypernetworks. The monotonicity is guaranteed by restricting the weight parameters of the mixing network output by the hypernetworks to be positive.

However, because current value decomposition methods, e.g. VDN, QMIX and QTRAN, only model the mean value of the joint state-action value, none of the existing methods take the random factors in the environment into consideration. So the baselines may have poor performance in some scenarios.

\section{MMD-MIX}
\begin{figure*}[htp]
    \centering
    \includegraphics[width = 6 in]{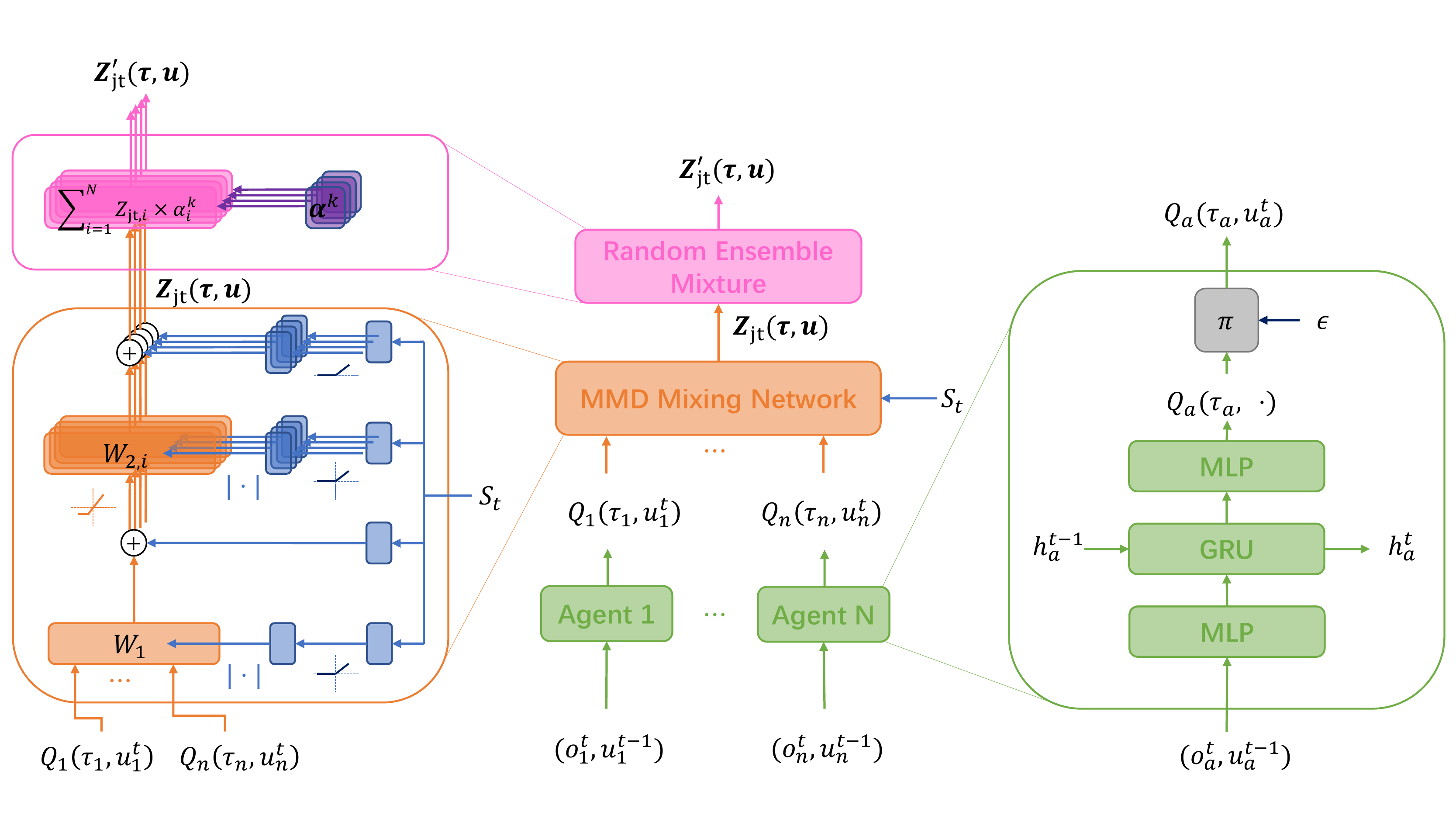}
    \caption{The overall architecture of MMD-MIX with REM. The right is agent $a$'s deep recurrent Q-network. The input of agent networks are the action-observation history record $\tau^a$ and they output the individual action-values. The left is the MMD mixing network, which mixes $Q_a(\tau_a,u_a^t)$ together with $s_t$ and outputs the joint state-action distribution $\left(Z_{\text{jt}, i}\right)_{i=1}^{ N}$ in the form of particles. Finally, a set of new particles $\left(Z_{\text{jt}, k}^\prime\right)_{k=1}^{K}$ are obtained through the REM module. In this figure, $N=K=4$.}
    \label{fig:framework}
\end{figure*}

\begin{table*}[htbp]
\centering
\label{table:scenario}
\caption{Maps in different scenarios.}
\begin{tabular}{lcccc}
\hline
Name&Ally Units&Enemy Units&Type&Difficulty\\
\hline
& & & &  \\[-6pt]
2s3z&\tabincell{c}{2 Stalkers\\3 Zealots}&\tabincell{c}{2 Stalkers\\3 Zealots}&\tabincell{c}{Heterogeneous, Symmetric}&Easy\\
\hline
& & & &  \\[-6pt]
3s5z&\tabincell{c}{3 Stalkers\\5 Zealots}&\tabincell{c}{3 Stalkers\\5 Zealots}&\tabincell{c}{Heterogeneous, Symmetric}&Easy\\
\hline
& & & &  \\[-6pt]
1c3s5z&\tabincell{c}{1 Colossus\\3 Stalkers\\5 Zealots}&\tabincell{c}{1 Colossus\\3 Stalkers\\5 Zealots}&\tabincell{c}{Heterogeneous, Symmetric}&Easy\\
\hline
& & & &  \\[-6pt]
2c\_vs\_64zg&2 Colossi&64 Zerglings&\tabincell{c}{Heterogeneous, Asymmetric,\\Large action space}&Hard \\
\hline
& & & &  \\[-6pt]
3s\_vs\_5z&3 Stalkers&5 Zealots&\tabincell{c}{Heterogeneous, Asymmetric}&Hard \\
\hline
& & & &  \\[-6pt]
27m\_vs\_30m&27 Marines&30 Marines&\tabincell{c}{Homogeneous, Asymmetric}&Super Hard \\
\hline
& & & &  \\[-6pt]
MMM2&\tabincell{c}{1 Medivac\\2 Marauders\\7 Marines}&\tabincell{c}{1 Medivac\\3 Marauder\\8 Marines}&\tabincell{c}{Heterogeneous, Asymmetric,\\Macro tactics}&Super Hard\\
\hline
\end{tabular}
\end{table*}

\begin{figure*}[ht]
    \centering
    \subfigure[2s3z]{
        \includegraphics[width=1.7 in]{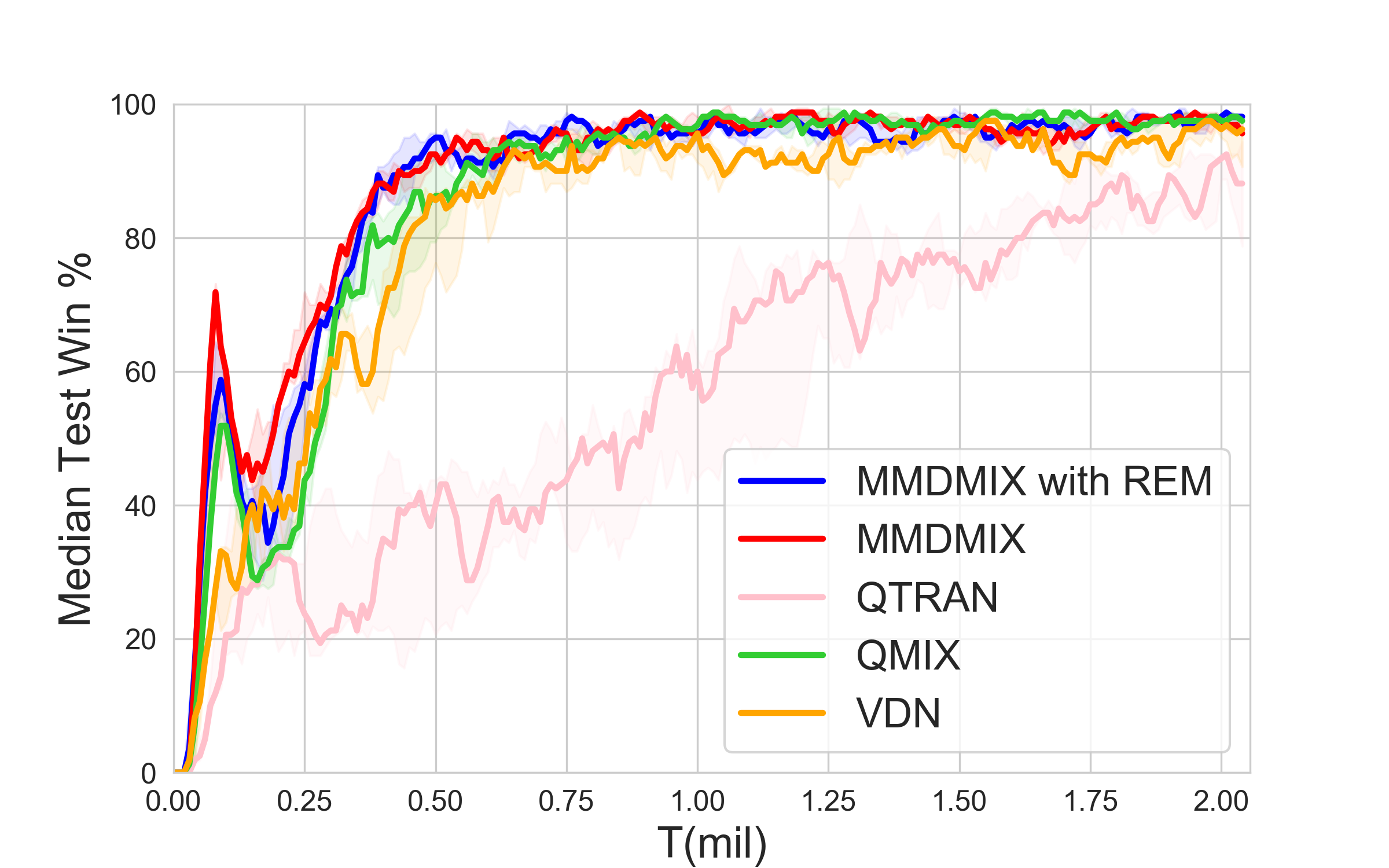}
    }
    \hspace{-0.3 in}
    \subfigure[3s5z]{
        \includegraphics[width=1.7 in]{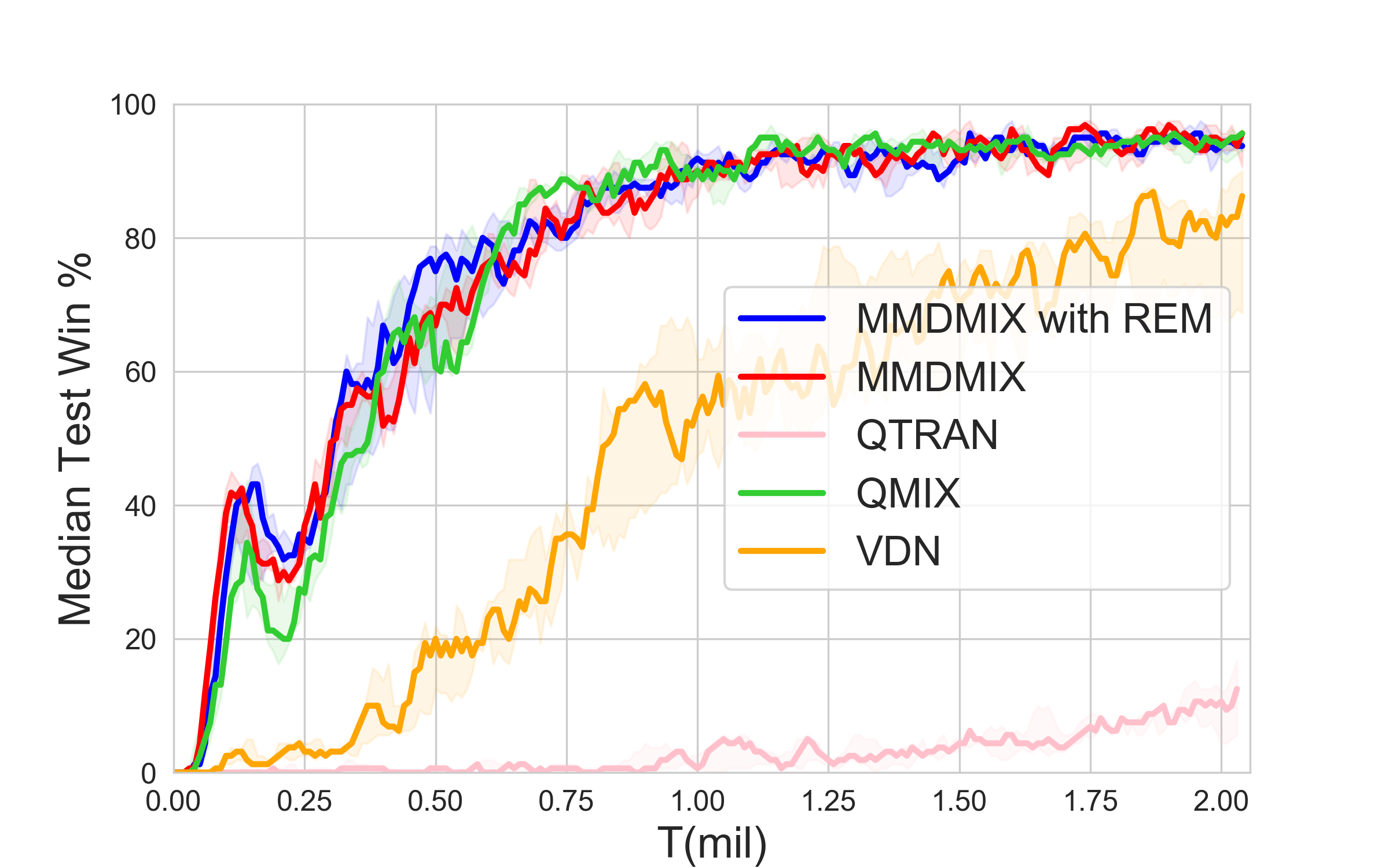}
    }
    \hspace{-0.3 in}
    \subfigure[1c3s5z]{
        \includegraphics[width=1.7 in]{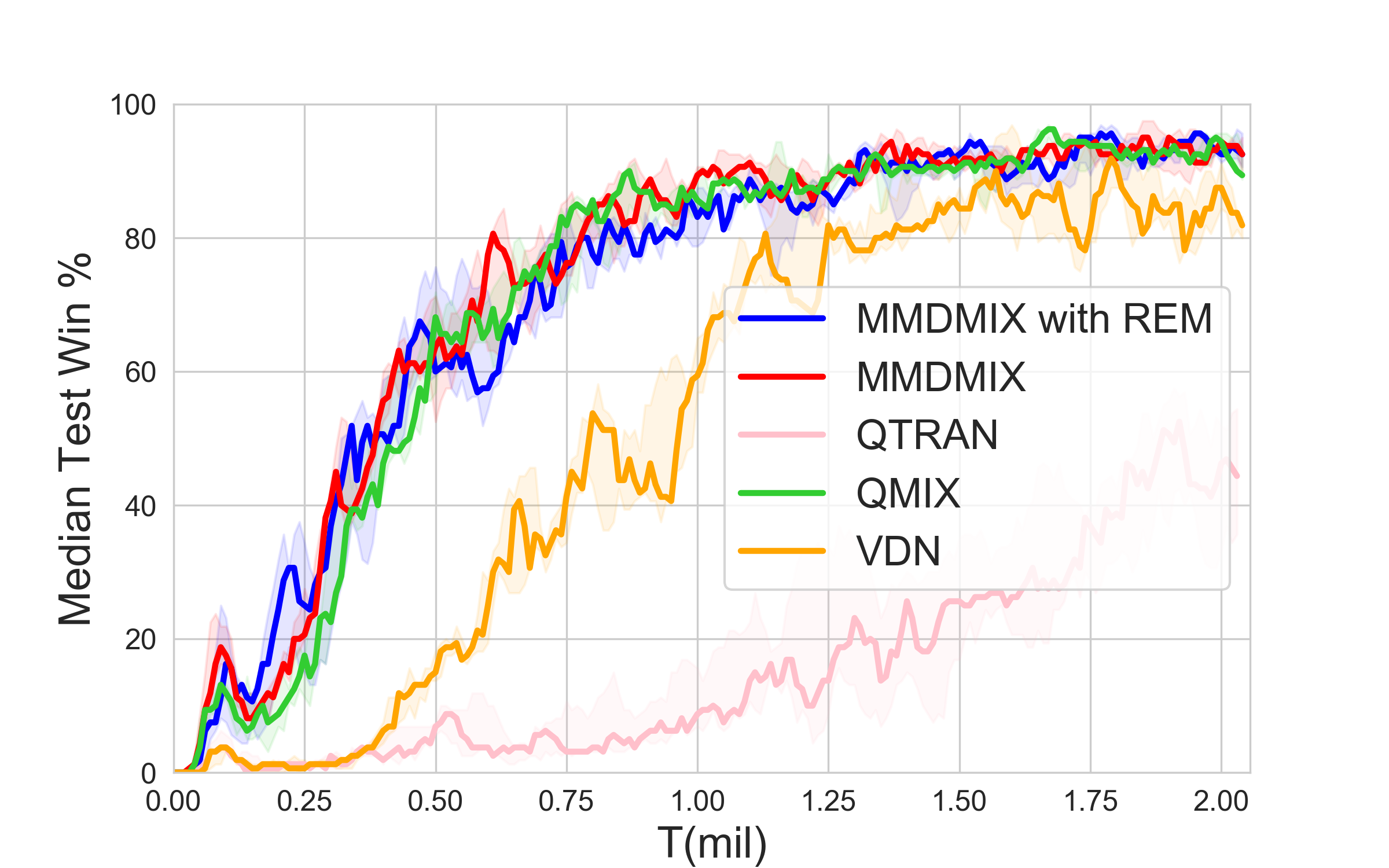}
    }

    \vspace{-0.1 in}
    \subfigure[2c\_vs\_64zg]{
        \includegraphics[width=1.7 in]{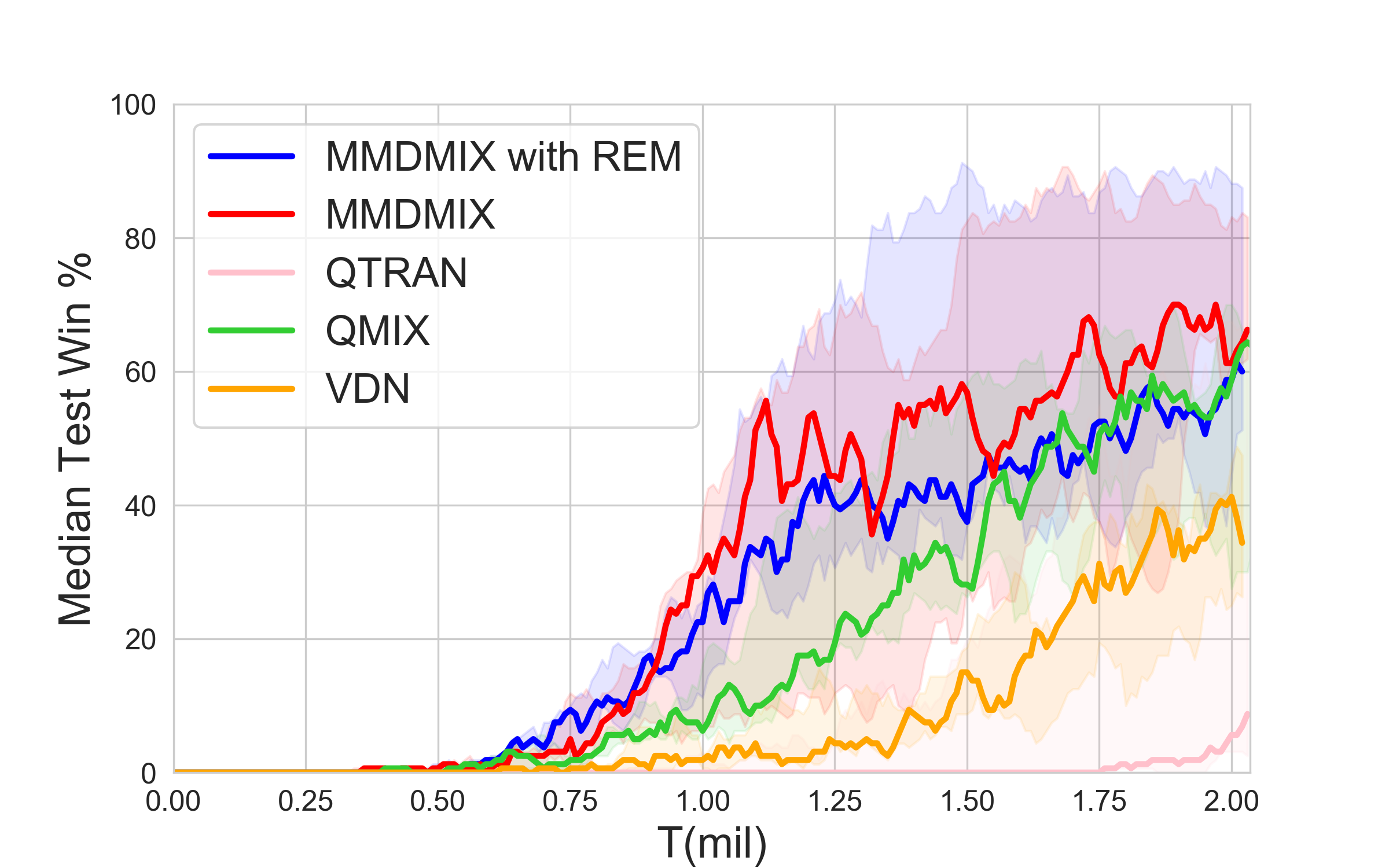}
    }
    \hspace{-0.3 in}
    \subfigure[3s\_vs\_5z]{
        \includegraphics[width=1.7 in]{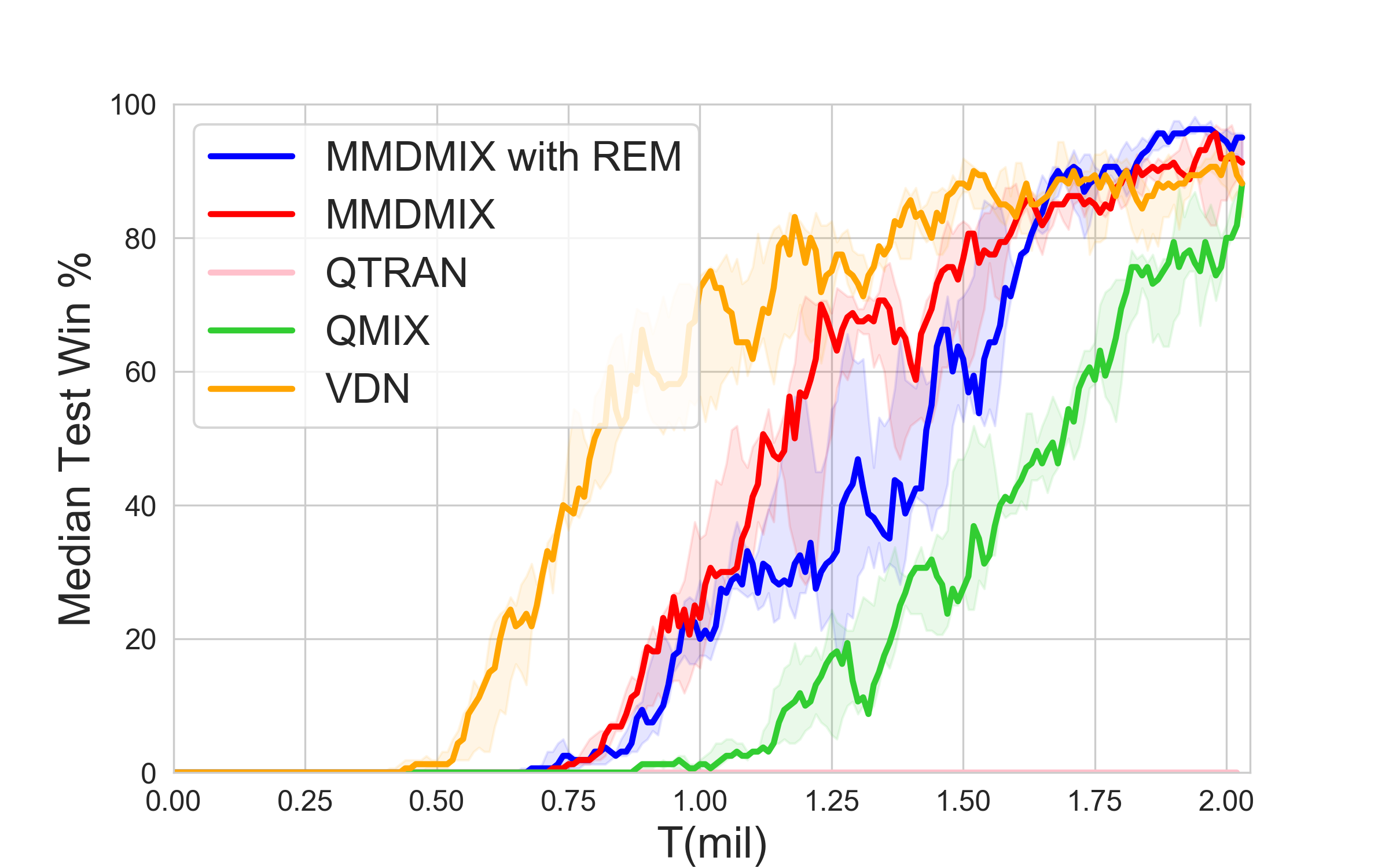}
    }
    \hspace{-0.3 in}
    \subfigure[27m\_vs\_30m]{
        \includegraphics[width=1.7 in]{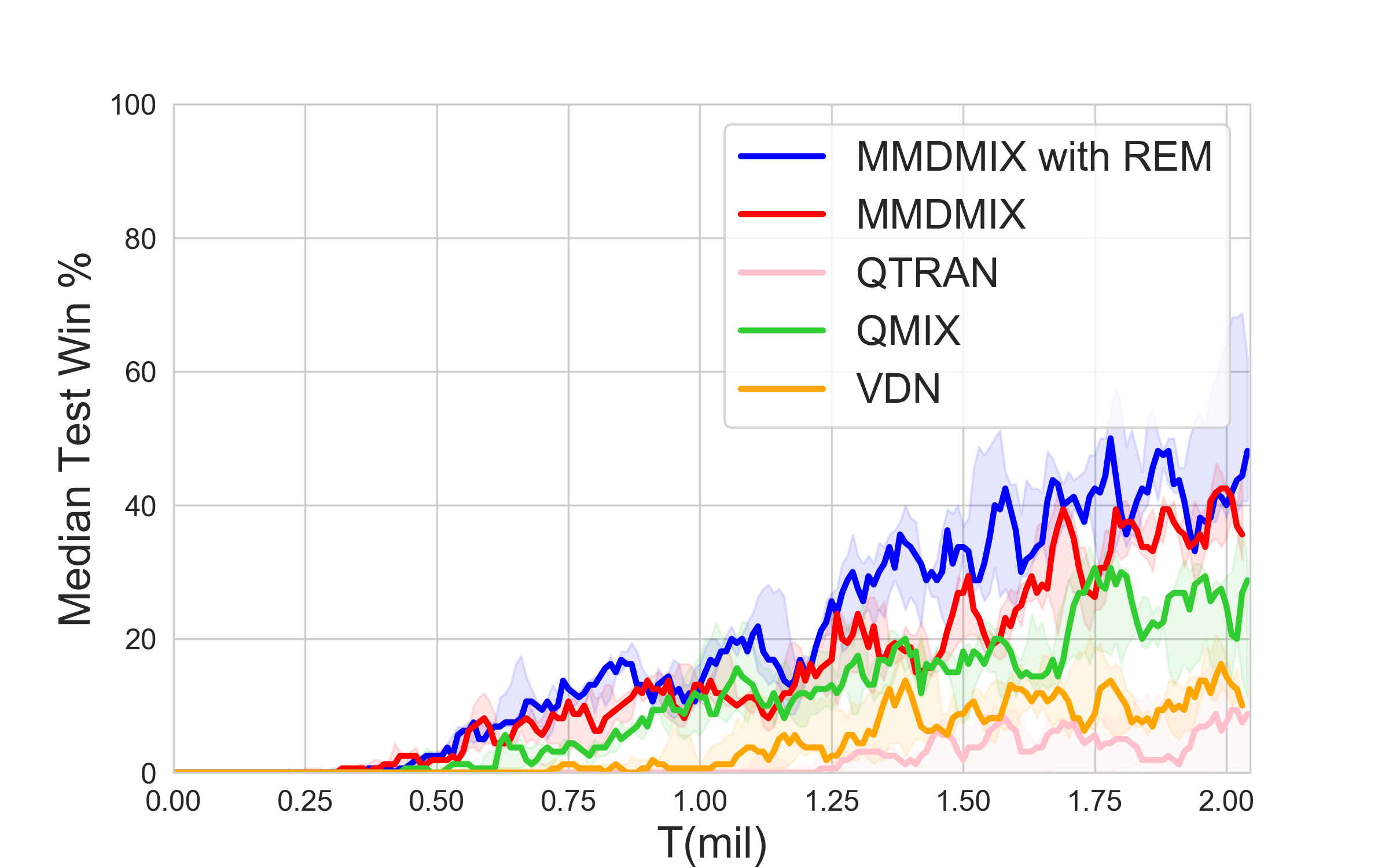}
    }
    \hspace{-0.3 in}
    \subfigure[MMM2]{
        \includegraphics[width= 1.7 in]{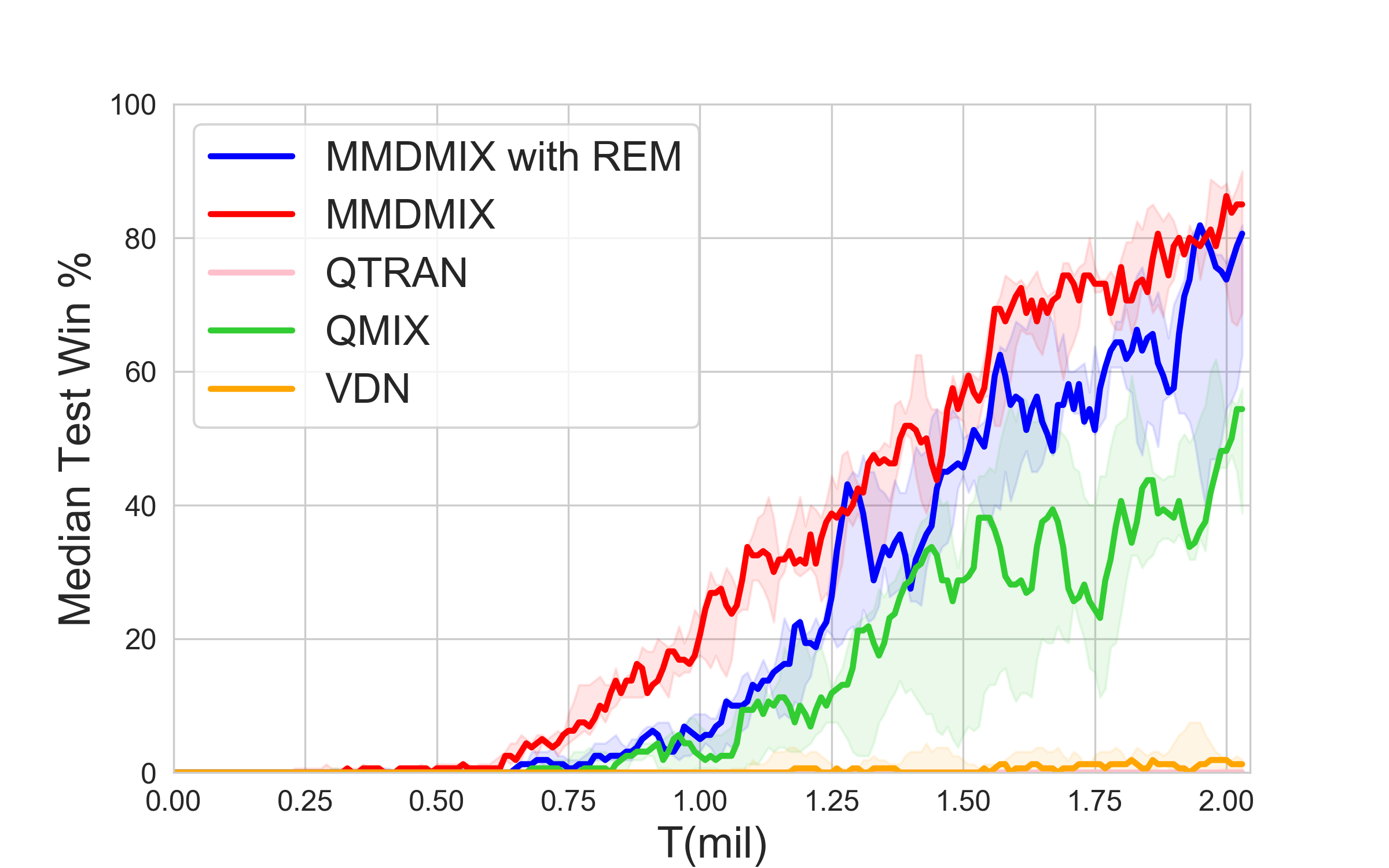}
    }
    \caption{Median win percentage of baselines and MMD-MIX on the easy scenarios(a-c), hard scenarios(d, e) and super hard scenarios(f, g).}
    \label{fig:result}
\end{figure*}

In this section, we propose a new method based on value decomposition called MMD-MIX. This method introduces the distributional reinforcement learning ideas in MMD-DQN into QMIX, so that the combined algorithm can effectively depicts the randomness of the environment. Besides, a reprocessing of the particles output by MMD-MIX is proposed. We think it can further improve the performance of MMD-MIX. 

\subsection{MMD Mixing Network}
MMD-DQN is constructed on the basis of the DQN structure. By changing the output size of the last layer to $N \times |\mathcal{U}|$ instead of $|\mathcal{U}|$, it represents the particle $\left(Z_{ \theta}(s, u)_{i}\right)_{i=1}^{N}$. A similar structure is also adopted in our proposed MMD-MIX, changing the original Mixing Network that only outputs one joint action-value in QMIX into an MMD Mixing Network that outputs multiple joint action-values. The specific method is to increase the number of hypernetworks that generate the weight parameters of the last layer of mixing network, and each group of weight parameters corresponds to a joint action value. In this way, for the individual action-value $Q_a$ of all agents, MMD-MIX will output a set of particles $\left(Z_{\text{jt},a}\right)_{a=1}^{ N}$ to represent the distribution of joint action-value.

In the same way as QMIX restricts the parameters of the Mixing Network, MMD-MIX can also use the absolute value of the parameters of the Mixing Network to ensure that the IGM conditions are established. So for each particle $Z_{\text{jt},\ j}$ output by MMD-MIX, all satisfy:
\begin{equation*}
    \frac{\partial \mathbb{E} Z_{\text{jt},i}(\boldsymbol{\tau}, \boldsymbol{u})}{\partial Q_{a}\left(\tau_{a}, u_{a}\right)} \geq 0, \;\; \forall a \in \{1, \dots, n\},  \;\; \forall i\in \{1,\dots,N\},
\end{equation*}
where $n$ represents the number of agents, and $N$ represents the number of particles.

\subsection{REM Module}

MMD Mixing Network outputs multiple joint action-values, which satisfies the structural requirements of REM. But because vanilla REM only generates one value after combining, some modifications are needed. We use $K$ categorical distributions $\alpha^k \sim \mathbf{P}_\Delta$, $k \in \{1,\dots,K\}$. In order to adapt to the $N$ outputs of MMD Mixing Network, the $\mathbf{P}_\Delta$ here is a uniform distribution Uniform (0, 1) over the standard ($N-1$)-simplex, i.e.,
\begin{equation*}
    \begin{aligned}
    \Delta^{\mathrm{N}-1}=\{\alpha^k \in \mathbb{R}^{\mathrm{N}}:  \alpha_{1}^k+\alpha_{2}^k+ \cdots +\alpha_{N}^k=1, \\
    \alpha_{i}^k \geq 0, \quad i=1, \ldots, N, \quad k=1,\dots,K \}.
    \end{aligned}
\end{equation*}
Then we get the new particles processed by REM:
\begin{equation}
    \label{eq:rem}
    Z_{\text{jt}, k}^{\prime}(\boldsymbol{\tau}, \boldsymbol{u}) = \sum_{i=1}^N \alpha^k_i Z_{\text{jt},i}(\boldsymbol{\tau}, \boldsymbol{u}).
\end{equation}
At the same time, since each element in $\boldsymbol{\alpha}$ is normalized, it is a non-negative value. This premise ensures that the monotonicity assumption can be obtained
\begin{equation*}
\centering
    \frac{\partial \mathbb{E} Z_{\text{jt},k}^{\prime}(\boldsymbol{\tau},\boldsymbol{u})}{\partial Q_{a}\left(\tau_{a}, u_{a}\right)}=\sum_{i=1}^N \alpha_i^k \frac{\partial \mathbb{E} Z_{\text{jt}, a}(\boldsymbol{\tau}, \boldsymbol{u})}{\partial Q_{a}\left(\tau_{a}, u_{a}\right)} \geq 0,
\end{equation*}
\begin{equation*}
        \\  \;\; \forall a \in \{1, \dots, n\},  \;\; \forall i\in \{1,\dots,N\},
    \;\; \forall k\in \{1,\dots,K\}.
\end{equation*}
This proof is an intuitive extension of MMD-MIX's monotonic constraint. We represent the REM algorithm as a module of the entire network structure, as shown in Figure~\ref{fig:framework}. Particles obtained through the MMD Mixing Network are processed by the REM module and we get a new set of particles finally.

Regarding convergence, since the REM module in this paper obtains multiple convex combination Q-value estimates instead of a single Q-value estimate, it is impossible to get all the Q-heads represent identical Q-functions and most importantly, all Q-heads will not converge according to the analysis mentioned in \cite{Agarwal2020AnOP}. So here we introduce a relatively strong assumption:

\noindent
\textbf{Proposition 1:} \emph{Consider the assumptions: (i) At any global minimum of $\mathcal{L}_{M M D^{2}}(\boldsymbol{\alpha},\theta)$, the Q-function heads $Z_{\text{jt},i}(\theta)$ for $i = 1,\dots,N$ minimize $\mathcal{L}_{M M D^{2}}(\boldsymbol{\alpha},\theta)$ for any $\alpha^k \in \Delta N - 1$, where $\theta$ are the parameters of the neural network and $\boldsymbol{\alpha} = [\alpha^1, \alpha^2, \dots, \alpha^K]$. (ii) The optimal distribution $Z_{\text{jt}}^\ast$ is defined in terms of the MDP induced by the data distribution and $Z_{\text{jt}}^\ast$ lies in the family that our functions can approximate. Then our proposed algorithm will converge to $Z_{\text{jt}}^\ast$. }

If the above assumptions are true, convergence is guaranteed. Although the assumption \emph{(i)} is difficult to meet, REM can still achieve good results in subsequent experiments.

\subsection{Loss Function}
Since there are multiple choices for the kernel function in the maximum mean discrepancy given by Equation (\ref{eq:mmd_distance}), the loss function for training MMD Mixing Network has many forms.  There are many common kernel function choices include RBF kernel, unrectified triangle kernel and so on.  For the reinforcement learning problem, as mentioned in \cite{Nguyen2020DistributionalRL}, only a kernel function that satisfies sum invariant and scale sensitive (such as triangle kernel) can guarantee that the Bellman operator $\mathcal{T}^\pi$ is a contraction operator, which is directly related to whether $\mathcal{T}^\pi$ can lead to convergence to a fixed point. However, some kernel functions which don't satisfy sum invariant and scale sensitive also have excellent performance in practice.

The current distribution can be expressed as $Z_{\text{jt}, k}^{\prime}(\boldsymbol{\tau},\boldsymbol{u})$, and Bellman target particles are as following formula:
\begin{equation*}
    \mathcal{T}Z_{\text{jt}, k}^{\prime}(\boldsymbol{\tau^\prime}, \boldsymbol{u^{\prime}})=r + \gamma Z_{\text{jt}, k}^{\prime}(\boldsymbol{\tau^\prime}, \boldsymbol{u^{\prime}}).
\end{equation*}
Then minimize the squared empirical MMD distance
\begin{equation}
\label{eq:new_loss}
\mathcal{L}_{M M D^{2}}\left(\left(Z_{\text{jt}, k}^{\prime}(\boldsymbol{\tau}, \boldsymbol{u})\right)_{k=1}^{K},\left(\mathcal{T}Z_{\text{jt}, k}^{\prime}(\boldsymbol{\tau^\prime}, \boldsymbol{u^{\prime}})\right)_{k=1}^{K} \right)
\end{equation}
given by Equation (\ref{eq:mmd_distance}) to update the parameters of the entire neural network. The full algorithm, which we call MMD-MIX with REM, is presented in Algorithm~\ref{algo:mmdmix}.
\begin{algorithm}
\label{algo:mmdmix}
\caption{MMD-MIX with REM}
\KwIn{Hyperparameters: $N$, $K$, $\gamma$, $\epsilon$}
\LinesNumbered
Initialize the replay buffer $D$\\
Initialize the parameters $\theta$ of the agent network and the mixing network\\
Initialize the target parameters $\theta^-$\\
\For{$episode \leftarrow$ 1 \textbf{to} $M$}
{Observe initial state $\boldsymbol{s}^1$ and observation $o^1_a$ for each agent $a$\\
    \For{$t \leftarrow 1$ \textbf{to} $T$}
    {
        \For{$a \leftarrow 1$ \textbf{to} $n$}
        {With probability $\epsilon$ select a random action $u_a^t$\\
        Otherwise $u_a^t=\arg\max_{u}Q_a(\tau^t_a, u)$}
    Take the joint action $\boldsymbol{u}^t$, and get the next observation $o^{t+1}_a$ for each agent and the reward $r^t$\\
    Store the transition $(s^t, \boldsymbol{o}^t, \boldsymbol{u}^t, r^t, s^{t+1}, \boldsymbol{o}^{t+1})$}
Sample a random mini-batch data from $D$\\
Sample $K$ categorical distributions $\alpha^k \sim \mathbf{P}_\Delta$, $k \in \{1,\dots,K\}$, where $\mathbf{P}_\Delta$ is a uniform distribution Uniform (0, 1) over the standard ($N-1$)-simplex\\
Calculate the particles $Z_{\text{jt},i}(\boldsymbol{\tau}, \boldsymbol{u})$ output by the MMD Mixing network\\
Calculate the particles $Z_{\text{jt}, k}^{\prime}(\boldsymbol{\tau}, \boldsymbol{u})$ output by REM module defined by Equation (\ref{eq:rem})\\
Calculate $\mathcal{L}_{M M D^{2}}$ defined by Equation (\ref{eq:new_loss})\\
Update $\theta$ by minimizing $\mathcal{L}_{M M D^{2}}$\\
Update target network parameters $\theta^-=\theta$ periodically}
\end{algorithm}

\section{Experiment}
In this section, we will evaluate the performance of the MMD-MIX algorithm in StarCraft II decentralized micromanagement tasks, and conduct an ablation experiment to further illustrate the role of REM.
\subsection{Settings}
We use StarCraft Multi-Agent Challenge (SMAC)\cite{samvelyan19smac} environment as our experimental platform. SMAC is based on the popular real-time strategy (RTS) game StarCraft II and it is a well-known experimental platform for Dec-POMDP problems. It contains a series of micromanagement scenarios. Each agent in SMAC can only observe partial state information and share a reward function. Through the performance of multi-agents in these different scenarios, the decision making ability of multi-agents in complex problems can be evaluated.

According to the difficulty of the task, SMAC divides the scenarios into three levels: easy, hard and super hard scenarios. We selected representative scenes for experiments in all three levels. The easy scenes include \emph{1c3s5z}, \emph{2s3z}, \emph{3s5z}, the hard scenes include \emph{2c\_vs\_64zg} and \emph{3s\_vs\_5z}, and the super hard scenes include \emph{MMM2} and \emph{27m\_vs\_30m}. These scenarios include challenges such as heterogeneous, asymmetric, large action space, and macro tactics as shown in Table~\ref{table:scenario}.

Our experiment is based on Pymarl\cite{samvelyan19smac}. We use QMIX and VDN in Pymarl as the baseline algorithm to compare with our proposed MMD-MIX. For QTRAN, We also test it on all of the scenarios mentioned above. QTRAN fails to achieve good performance and we think because its practical relaxations could impact the accuracy of its updating. We run each experiment independently 5 times, and each independent run takes between 5 to 14 hours, depending on the exact scenario, using Nvidia GeForce RTX 3090 graphics cards and Intel(R) Xeon(R) Platinum 8280 CPU. We use the median of win ratio instead of the mean in order to avoid the effect of any outliers. The evaluation progress can be estimated by periodically running a fixed number of evaluation episodes (in practice, 32) without any exploratory behaviours. The version of our Starcraft II is 4.6.2(B69232) as same as \cite{samvelyan19smac}.

All hyperparameters in MMD-MIX are the same as the hyperparameters in QMIX and VDN algorithms in Pymarl. The hypernetworks producing the both weights of the mixing network consist of a single hidden layer of 64 units with a ReLU non-linearity and the one producing the final bias of the mixing network consists of a single hidden layer of 32 units. We set all neural networks to be trained by using RMSprop with learning rate 0.0005. In addition, for the unique hyperparameters in MMD-MIX, the number of particles $N$, through experimental comparison we set $N=8$ to achieve the balance of performance to computational overhead. Out of consideration to avoid the influence of other factors, we set the number of categorical distributions to 8, i.e., $K = N = 8$. Comparing the experimental results, we set the kernel function in the loss function $\mathcal{L}_{MMD^{2}}$ to unrectified triangle kernel $k(x,y)=-\| x-y \|^2$ which satisfies sum invariant and scale sensitive. The advantage of this setting is that the number of training hyperparameters can be reduced and it can ensure that the MMD distance can lead to better results when there are few particles. 
\subsection{Validation}
\begin{table}[htbp]
\centering
\caption{Median performance of the test win percentage (\%) in different scenarios.}
\begin{tabular}{lccccc}
\hline
Scenario&\tabincell{c}{MMD-MIX \\ with REM}&MMD-MIX&QTRAN&QMIX&VDN\\
\hline
& & & & & \\[-6pt]
2s3z&\textbf{99}&\textbf{99}&93&98&98 \\
3s5z&96&\textbf{97}&13&96&87 \\
1c3s5z&\textbf{96}&95&47&95&88\\
2c\_vs\_64zg&61&\textbf{70}&9&64&41\\
3s\_vs\_5z&\textbf{96}&\textbf{96}&0&88&93 \\
27m\_vs\_30m&\textbf{48}&42&9&29&16\\
MMM2&82&\textbf{86}&0&62&1\\
\hline
\end{tabular}
\label{table:result_table}
\end{table}
Figure~\ref{fig:result} shows the performance of our proposed MMD-MIX algorithm and baseline algorithms in different scenarios. Among them, in order to study the impact of REM module in the algorithm, we use pure MMD-MIX and MMD-MIX with REM to compare. Each solid line in Figure~\ref{fig:result} represents the median win ratio and 25\%-75\% percentile is shaded. It can be found that except in the scenario \emph{3s\_vs\_5z}, the learning speed of MMD-MIX is faster than other algorithms in other scenarioss. This phenomenon becomes more clear as the difficulty level of the scenario increases.

Table~\ref{table:result_table} shows the median test win rate of different algorithms. It can be seen that MMD-MIX is better than the baseline algorithms in different scenarios, especially in hard and super hard scenarios.

From the ablation experiment for REM, it can be found that REM plays a small role in easy scenarios, and even slows down the learning speed to a certain extent in a few scenarios. But in some scenarios with higher levels of difficulty, especially the scenario \emph{27m\_vs\_30m}, it can be seen that the exploration ability brought by REM can significantly improve the learning speed of the algorithm. In addition, in the map \emph{3s\_vs\_5z}, even though the learning speed of MMD-MIX with REM is relatively slow in the first half, the learning effect in the second half exceeds all other algorithms. We believe that the reason for this phenomenon is that only in more difficult scenarios, the exploration ability brought by the random noise introduced by REM can really play a significant part.
\section{Conclusion and Future work}
In this paper, we propose a simple yet effective method of combining distributional reinforcement learning and random ensemble mixture, and apply it to multi-agent reinforcement learning. First, we introduce MMD-MIX, a multi-agent collaborative algorithm that introduces Maximum Mean Discrepancy. Meanwhile, in order to explicitly introduce randomness and improve the exploration ability of the algorithm, we utilize the REM module to optimize the effect of the algorithm. Experiments show that the performance of our proposed algorithm is significantly improved compared to the baseline algorithms, especially in hard and super hard scenarios.

Some distributional reinforcement learning algorithms, such as IQN and FQF, all generate Q-value distributions in a similar way to DGMs (Deep Generative Models)\cite{Kingma2014AutoEncodingVB, Goodfellow2014GenerativeAN} and get good results. Therefore, transforming MMD Mixing Network into DGMs form is one of our future works. In addition, REM is often used in off-line cases, and it is also an interesting direction to improve the performance of the algorithm in off-line scenarios.

\bibliography{main.bbl}
\bibliographystyle{IEEEtran}

\end{document}